# Hyperbolic and Circular Trigonometry and Application to Special Relativity

G. Dattoli[a)] and M. Del Franco [b)]

ENEA - FIS-MAT, Centro Ricerche Frascati,

C.P. 65 - 00044 Frascati, Rome (Italy)

b) ENEA Guest

**Abstract**

We discuss the most elementary properties of the hyperbolic trigonometry and show how they can be exploited to get a simple, albeit interesting, geometrical interpretation of the special relativity. It yields indeed a straightforword understanding of the Lorentz transformation and of the relativistic kinematics as well. The geometrical framework adopted in the article is useful to disclose a wealth of alternative *trigonometries* not taught in undergraduate and graduate courses. Their introduction could provide an interesting and useful conceptual tool for students and teachers.

a) Corresponding author:
   G. Dattoli
   Tel +39 06 94005421
   e-mail Dattoli@frascati.enea.it
   Fax +39 06 94005334

**I. INTRODUCTION**

The geometrical interpretation of the circular functions is well known, but that of their hyperbolic counterparts does not appear equally well known as it should be, even among very well experienced scientists. The importance of the hyperbolic trigonometry for the study of special relativity does not appear widspread known as it should be even though it may provide a useful pedagocial tool. This article is devoted to an introduction to this subject with some applications to special relativity, with particular reference to Lorentz transformations and to kinematics.

In Figures 1 we have reported the hyperbolic functions defined on arcs of a hyperbola, which plays the same role of the unit circle in the case of ordinary trigonometry.

The hyperbolic sine and cosine functions satisfy indeed the identity

$$\cosh(\alpha)^2 - \sinh(\alpha)^2 = 1 \tag{1}$$

and the argument $\alpha$ which is the counterpart of the angle in ordinary trigonometry, is given by the area sector, dashed in Fig. 1a, ref. 1.

In Figure 1b we have reported the geometrical interpretation of the hyperbolic secant and tangent.

A bridge between Circular and Hyperbolic functions can be realized using either the imaginary numbers or the geometrical construction reported in Fig. 2 and based on the identity

$$\operatorname{sech}(\alpha)^2 + \tanh(\alpha)^2 = 1 \tag{2}$$

The definition of the Gudermann function $\Phi(\alpha)$ follows, almost naturally, from the above identity and it can be specified through the following relations

$$\begin{aligned}\cos(\Phi(\alpha)) &= \operatorname{sech}(\alpha), \\ \sin(\Phi(\alpha)) &= \tanh(\alpha)\end{aligned} \tag{3}$$

as

$$\Phi(\alpha) = \tan^{-1}(\sinh(\alpha)) \tag{4}$$

In other words the function $\Phi(\alpha)$ is the angle shown in Fig. 2 expressed as a function of the sector area $\alpha$. This function has been reported in Fig. 3 along with the Hyperbolic secant and tangent and it is evident that $\Phi(\alpha)$ ranges from $-\frac{\pi}{2}$ to $\frac{\pi}{2}$.

In special relativity time and space coordinates are linked by the Lorentz invariant or Minkowsky norm (we limit ourselves to the time like case)

$$\tau^2 - X^2 = 1 \tag{5}$$

On account of eq. (1) the natural geometrical framework to study the space time properties linked by (5) is the hyperbolic trigonometry and therefore we set

$$\begin{aligned} X &= \sinh(\varphi), \\ \tau &= \cosh(\varphi) \end{aligned} \tag{6}$$

Furthermore, by defining the constant

$$\frac{X}{\tau} = \beta \tag{7}$$

Which can be reinterpreted as the relativistic reduced velocity [b], we find

$$1 - \left(\frac{X}{\tau}\right)^2 = 1 - \beta^2 = \frac{1}{\cosh(\varphi)^2} \tag{8}$$

Which can be exploited to derive the further identities

---

[b] It is to be noted that such an assumption is not necessary at all, we can define such a quantity independently of any physical interpretation.

$$\text{sech}(\varphi) = \gamma^{-1},$$
$$\tanh(\phi) = \beta,$$
$$\gamma = \frac{1}{\sqrt{1-\beta^2}}$$
(9)

and the relevant geometrical interpretation is given in Fig. 4, which can also be exploited to stress the following remarks:

a) A point representative of the Minkowsky chronotope can be defined on a circle and on an arc of hyperbola as well, they will be called reciprocal images

b) The radius of the circle is the Lorentz invariant [c]

c) When $\beta \to 1$ the reciprocal image points of the Minkowsky space can be visualized as moving either on a arc of circle and on arc of hyperbola; with the increasing of $\beta$ the hyperbolic secant decreases and the hyperbolic cosine increases

d) $\beta$ can never exceed unit, it is on one side the vertical projection of a point moving on a circle of unit radius and on the other side, it is associated with the hyperbolic tangent which cannot overcome the asymptote of the hyperbola.

Regarding this last point we note that when the reduced velocity approaches 1, the point $P$ moves on the positive branch of the hyperbola and on the positive quarter of circle of the first quadrant.

After these introductory remarks we will discuss some consequences of Physical nature.

## II. THE LORENTZ TRANSFORMATION AND THE HYPERBOLIC TRIGONOMETRY

Any transformation preserving the norm given in eq. (5) is a "rotation" in the Minkowsky space realized as

$$\begin{pmatrix} \tau' \\ X' \end{pmatrix} = \hat{R}_h(\varphi_1) \begin{pmatrix} \tau \\ X \end{pmatrix},$$

$$\hat{R}_h(\varphi) = e^{\varphi \hat{h}} = \begin{pmatrix} \cosh(\varphi) & \sinh(\varphi) \\ \sinh(\varphi) & \cosh(\varphi) \end{pmatrix} \qquad (10)$$

$$\hat{h} = \begin{pmatrix} 0 & 1 \\ 1 & 0 \end{pmatrix}$$

---

[c] The norm of the space defined on the circle is Euclidean.

Which is immediately recognized as a Lorentz transformation, since from eq. (9) it follows that

$$\begin{aligned} \gamma &= \cosh(\varphi), \\ \beta \gamma &= \sinh(\varphi) \end{aligned} \qquad (11a)$$

Two successive Lorentz transformation are therefore understood as the matrix product

$$\hat{R}_h(\varphi_1) \hat{R}_h(\varphi_2) = e^{(\varphi_1 + \varphi_2) \begin{pmatrix} 0 & 1 \\ 1 & 0 \end{pmatrix}} = \hat{R}_h(\varphi_1 + \varphi_2) \qquad (11b)$$

Regarding the transformation of $\beta$ we obtain therefore

$$\begin{aligned} \beta' &= \tanh(\varphi_1 + \varphi_2) = \frac{\sinh(\varphi_1)\cosh(\varphi_2) + \sinh(\varphi_2)\cosh(\varphi_1)}{\cosh(\varphi_1)\cosh(\varphi_2) + \sinh(\varphi_1)\sinh(\varphi_2)} = \\ &= \frac{\beta_1 + \beta_2}{1 + \beta_1 \beta_2} \end{aligned} \qquad (12)$$

Which can be viewed as the relativistic composition of the velocities, emerging within the present context, as a pure geometrical property.

In Figure 5 we have reported the geometrical meaning of the composition, while the physical meaning in terms of time dilatation will be discussed in the concluding section.

It is furthermore worth noting that

$$\begin{aligned} \gamma' &= \cosh(\varphi_1 + \varphi_2) = \cosh(\varphi_1)\cosh(\varphi_2) + \sinh(\varphi_1)\sinh(\varphi_2), \\ &= \gamma_1 \gamma_2 (1 + \beta_1 \beta_2) \end{aligned} \qquad (13a)$$

and

$$(\gamma \beta)' = \sinh(\varphi_1 + \varphi_2) = \sinh(\varphi_1)\cosh(\varphi_2) + \sinh(\varphi_2)\cosh(\varphi_2) =$$
$$= \gamma_1\gamma_2\beta_2 + \gamma_2\gamma_1\beta_2 = \gamma_1\gamma_2(\beta_1 + \beta_2) \quad (13\ b)$$

The above identities can be rewritten as follows

$$\gamma' = \gamma_1[\gamma_2 + \beta_1(\beta_2\gamma_2)],$$
$$(\gamma\beta)' = \gamma_1[(\beta_2\gamma_2) + \beta_1(\gamma_2)] \quad (13\ c)$$

an can therefore be understood as the Lorentz transformation for energy and momentum since the following correspondence holds

$$\gamma \to E,$$
$$\beta\gamma \to p \quad (14)$$

The same result follows by considering the invariant

$$E^2 - (pc)^2 = (m_0 c^2)^2 \quad (15a)$$

and by setting

$$\frac{E}{m_0 c^2} = \cosh(\varphi) = \gamma,$$
$$\frac{p}{m_0 c} = \sinh(\varphi) = \beta\gamma \quad (15b)$$

We can therefore utilize the same geometrical point of view discussed so far, an analogous treatment was developed in ref. 2 by Foldy and Wouthuysen who introduced the triangle shown in Fig. 6 within the theory for the decoupling of small and large components of the Dirac equation.

Equation (14b) are defined by a direct inclusion of the rest mass, therefore their use for the case of massless particles may be doubtful.

For this reason we can take advantage from the identities reported in ref. 3

$$\tanh(\vartheta) = \frac{pc}{E+mc^2} = \frac{\gamma\beta}{\gamma+1}$$

$$\text{sech}(\vartheta) = \sqrt{\frac{2mc^2}{E+mc^2}} = \sqrt{\frac{2}{\gamma+1}}$$

(15 c)

They are framed within the geometrical picture developed in this paper as shown in Fig. (7a), where we have linked different relativistic kinematic quantities including the kinetic energy $T = E - m_0c^2$.

The above picture (Fig. 7b) allows a few interesting speculations, reported below

a. Massless particles in this picture are characterized by the point indicated with a small dot in Fig. 7b

b. The transition to massive particles occurs through a rotation by an angle specified by

(16) $$\Omega = \tan^{-1}\frac{p}{\sqrt{2m_0(E+m_0c^2)}}$$

c. Particles with negative masses and energies move on the negative branches of the circle and of the hyperbola.

It is important to realize that the transition from massless to massive particles through the angle given in eq. (16) cannot be considered strictly speaking a rotation. A rotation would indeed imply the equivalence between particles with and without mass. Such a rotation is however only formal, the circle is not symmetric, because the symmetry is actually naturally *broken* by the assumption that the allowed physical region is that reported in Fig. 8.

When the reciprocal points move on the half right plane, they describes either an angle (on the circle) and a sector area, (on the hyperbola). The angle is constrained to the interval (–π/2; π/2).

A more complete analysis should include the extension of the previous considerations to space-like vectors, but it is not reported here for reasons of space.

## III. CIRCULAR VS HYPERBOLIC

In the previous sections we have considered a kind of duality between hyperbolic and circular trigonometry. Here we will discuss the problem from the point of view of the ordinary trigonometry.

In Figure 9 we report the ordinary trigonometric circle and its *"hyperbolic"* counterpart realized using the identity

$$\sec(\alpha)^2 - \frac{1}{\tan(\alpha)^2} = 1 \tag{17}$$

Just to give a fairly direct idea of such a duality, we can apply the picture discussed so far to the trajectories of the harmonic oscillator, whose Hamiltonian can be written as

$$E = \frac{1}{2m}\left(p^2 + k\,x^2\right) \tag{18}$$

If we define the variables

$$\begin{aligned}
\frac{p}{\sqrt{2mE}} &= \cos(\Omega t + \varphi), \\
\frac{m\Omega x}{\sqrt{2E}} &= \sin(\Omega t + \varphi), \\
\frac{m\Omega x}{p} &= \tan(\Omega t + \varphi), \\
\sqrt{\frac{E}{T}} &= \sec(\Omega t + \varphi), \\
T &= \frac{p^2}{2m}, \\
\Omega &= \sqrt{\frac{k}{m}}
\end{aligned} \tag{19}$$

We can express the harmonic motion as shown in Fig. 10, we have in this case a completely arbitrary phase and any points on the circle (and on the arcs of the hyperbola) are equivalent.

Let us furthermore note that any Euclidean rotation can be generated as

$$\hat{R}_c(\alpha) = e^{\alpha \hat{i}} = \begin{pmatrix} \cos(\alpha) & \sin(\alpha) \\ -\sin(\alpha) & \cos(\alpha) \end{pmatrix},$$
$$\hat{i} = \begin{pmatrix} 0 & 1 \\ -1 & 0 \end{pmatrix} \quad (20)$$

where $\hat{i}$ is the *imaginary* unit matrix satisfying the identity

$$\hat{i}^2 = -\hat{1},$$
$$\hat{1} = \begin{pmatrix} 1 & 0 \\ 0 & 1 \end{pmatrix} \quad (21)$$

The exponential matrix, appearing in eq. (20), can be decomposed as the ordered product of three matrices as it follows [4]

$$e^{\alpha \hat{i}} = e^{g_+(\alpha)\hat{\sigma}_+} e^{g_0(\alpha)\hat{\sigma}_0} e^{g_-(\alpha)\hat{\sigma}_-} \quad (22)$$

where

$$\hat{\sigma}_+ = \begin{pmatrix} 0 & 1 \\ 0 & 0 \end{pmatrix}, \quad \hat{\sigma}_- = \begin{pmatrix} 0 & 0 \\ 1 & 0 \end{pmatrix},$$
$$\hat{\sigma}_0 = \begin{pmatrix} 1 & 0 \\ 0 & -1 \end{pmatrix} \quad (23)$$

are the Pauli matrices.

The ordering functions $g_{\pm,0}(\alpha)$ are easily obtained from eq. (22) and reads

$$g_+(\alpha) = -g_-(\alpha) = \tan(\alpha)$$
$$e^{g_0(\alpha)} = \sec(\alpha) \quad (24)$$

Which are reported in Fig. 11 along with their geometrical interpretation.

## IV. CONCLUDING REMARKS

Let us now consider a body of mass *m* moving under the action of a constant external force *F*, the equation of motion yielding the time dependence of the velocity can be written as

$$\frac{d}{dt}(\gamma v) = a,$$
$$a = \frac{F}{m} \tag{25}$$

which is straightforwardly integrated thus getting

$$\beta = \frac{1}{c}\frac{at}{\sqrt{1+\left(\frac{at}{c}\right)^2}},$$
$$\gamma = \sqrt{1+\left(\frac{at}{c}\right)^2} \tag{26}$$

or what is the same

$$\tanh(\varphi(t)) = \frac{1}{c}\frac{at}{\sqrt{1+\left(\frac{at}{c}\right)^2}},$$
$$\mathrm{sech}(\varphi(t)) = \frac{1}{\sqrt{1+\left(\frac{at}{c}\right)^2}} \tag{27}$$

Suppose now that we have used the above equations to describe the motion of a rocket and the time t is the earth time, the rocket time will therefore be given by

$$d\tau = \frac{dt}{\gamma} = \mathrm{sech}(\varphi(t))\,dt \tag{28}$$

This relation clarifies the role played by the segment OP in Fig. 4 as the time contraction or more properly it represents the Lorentz contraction [d].

Furthermore the Gudermann function is just given by

$$\Phi(t) = \tan^{-1}\left(\frac{at}{c}\right) \tag{29a}$$

while for the sector area we get

$$\varphi(t) = \ln\left[\frac{at}{c} + \sqrt{1 + \left(\frac{at}{c}\right)^2}\right] \tag{29b}$$

Supposing that the external force acting on the body is any generic time dependent force F(t), we find therefore that the angle $\Phi(t)$ can be written as

$$\Phi(t) = \tan^{-1}\left[\frac{\int_0^t F(t')\,dt'}{mc}\right] . \tag{29c}$$

We will now extend the previous geometrical picture to the description of elliptic functions. We believe that this topic too is not treated as it should in freshman courses. Very often Physics students are scared by this topic, which is treated in too abstract terms. To this aim we will adopt an elementary point of view to the theory of elliptic Jacobi functions[5], originally considered as inversion of the elliptic integrals (hence the name). Here we will follow a more elementary, but we believe more effective description, by treating them as trigonometric function on an ellipse.

In Figure 12 we have reported an ellipse, whose equation writes

---

[d)] Note that the racket "sees" the distance reduced.

$$\left(\frac{x}{a}\right)^2 + y^2 = 1, \qquad (30)$$
$$x^2 + y^2 = r^2$$

We define the quantity

$$u = \int_P^Q r \, d\vartheta \qquad (31)$$

and define the Jacobi elliptic functions as

$$\begin{aligned} y &= sn(u;k), \\ \frac{x}{a} &= cn(u;k), \\ r &= dn(u;k), \\ k &= \sqrt{1 - \frac{1}{a^2}} \end{aligned} \qquad (32)$$

Which can be combined with eqs. (31) to get the following fundamental relations

$$\begin{aligned} sn^2(u;k) + cn^2(u;k) &= 1, \\ dn^2(u;k) + k^2 sn^2(u;k) &= 1 \end{aligned} \qquad (33)$$

In Figure 12 we have also reported the analogous of the sec and tan functions defined below

$$\begin{aligned} nc(u;k) &= \frac{1}{cn(u;k)} \\ sc(u;k) &= \frac{sn(u;k)}{cn(u;k)} \end{aligned} \qquad (34)$$

in Figure 13 we have reported the transition from the Jacobi elliptic functions to their hyperbolic counterparts, which are usually not reported in the literature, since they can be defined as Jacobi functions with imaginary variables.

Acoding to the previous discussion it appears evident that circular, hyperbolic and elliptic trigonometries are just examples of the wealth of trigonometries one can define using elementary geometrical means, just to give a further example in figures 14 we report parabolic sine and cosine functions which are defined on an arc of parabola.

They satisfy the identity

$$\cos_{p_+}(\varphi)^2 + \sin_{p_+}(\varphi) = 1 \tag{35}$$

the relevant derivatives can be shown to be [6]

$$\frac{d}{d\varphi}\cos_{p_+}(\varphi) = -\frac{1}{1+\cos_{p_+}(\varphi)^2},$$
$$\frac{d}{d\varphi}\sin_{p_+}(\varphi) = 2\frac{\cos_{p_+}(\varphi)}{1+\cos_{p_+}(\varphi)^2} \tag{36}$$

The parabolic tangent and secant can be defined as shown in Fig. 14 and therefore from eq. (35) we find

$$\sec_{p_+}(\varphi)^2 - \sec_{p_+}(\varphi)\tan_{p_+}(\varphi) = 1 \tag{37}$$

The parabolic trigonometric functions are not transcendent but irrational functions and can be exploited as the natural solutions of problems involving third degree algebraic equations, or differential equations of the type

$$y' = -\frac{1}{1+y^2} \tag{38}$$

Albeit this article yield just a cursory look at the problems underlying *non standard* trigonometry and their applications, it perhaps offers the fueling of their usefulness as conceptual and pedagogical tool.

We have found that the introduction of a different point of view of teaching trigonometric concepts is extremely useful for Physics students, who seem to appreciate the concreteness of this tool, which yields a more general conceptual framework to understand Physical problems.

**ACKNOWLEDGMENTS**

The kind interest and encouragements of Dr. A. Doria are gratefully acknowledged.

**Figure Captions**

Fig. 1 -   a) definition of sinh($\alpha$) and cosh($\alpha$); b) definition of sech($\alpha$) and tanh($\alpha$)

Fig. 2 -   Hyperbolic vs. circular trigonometry

Fig. 3 -   Gudermann function $\Phi(\alpha)$ (red), hyperbolic tangent and hyperbolic secant vs.

Fig. 4 -   Geometrical interpretation of the Kinematic factors of the special relativity

Fig. 5 -   Geometrical meaning of the composition for the hyperbolic functions

Fig. 6 -   Foldy and Wouthuysen

Fig. 7 –   a) The geometry of the kinematics of the Special Relativity; b) Transition from mass less to massive particles

Fig. 8 -   Gudermann function vs. sector area

Fig. 9 -   The trigonometric circle, its hyperbolic counterpart and geometrical interpretation of the circular functions

Fig. 10 -  Armonic motion of P on the hyperbola for an arbitrary phase and any points on the circle

Fig.11 -   Characteristic functions $g(\alpha)$ and $e^{h(\alpha)}$

Fig. 12 -  Ellipse and definition of the Jacobi elliptic functions

Fig. 13 -  Jacobi elliptic functions

Fig. 14 -  Definition of the parabolic trigonometric functions

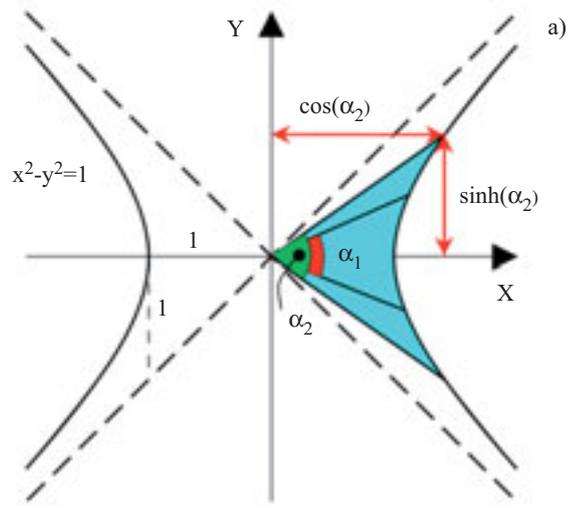

Fig. 1a)

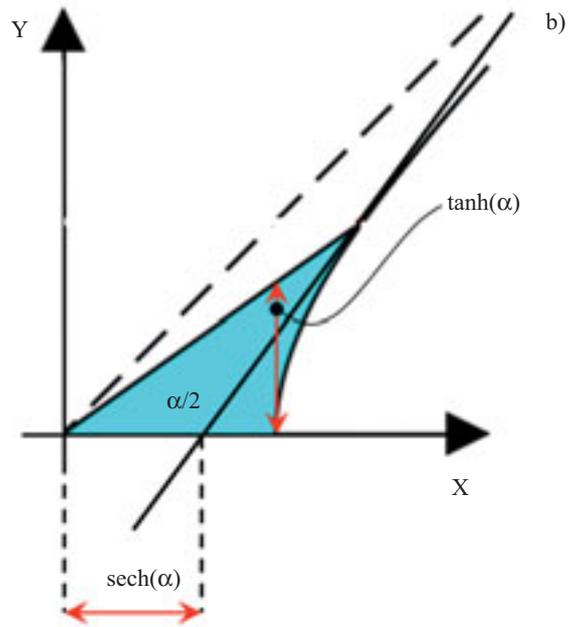

Fia. 1b)

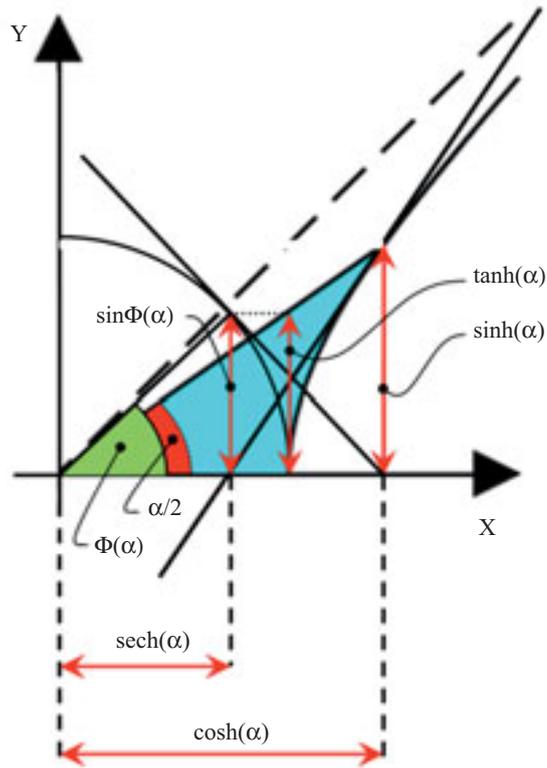

Fig. 2

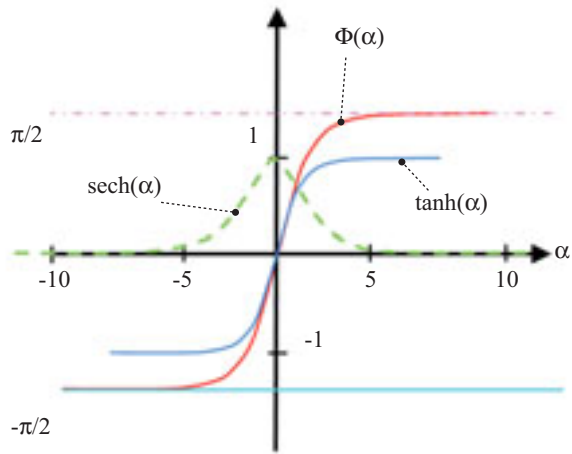

Fig. 3

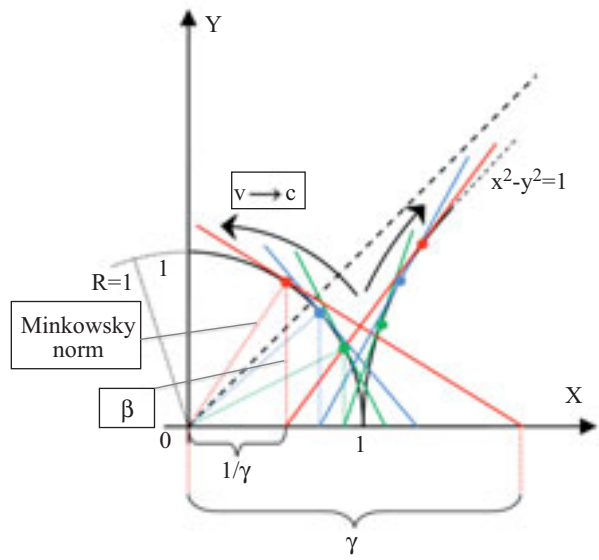

Fig. 4

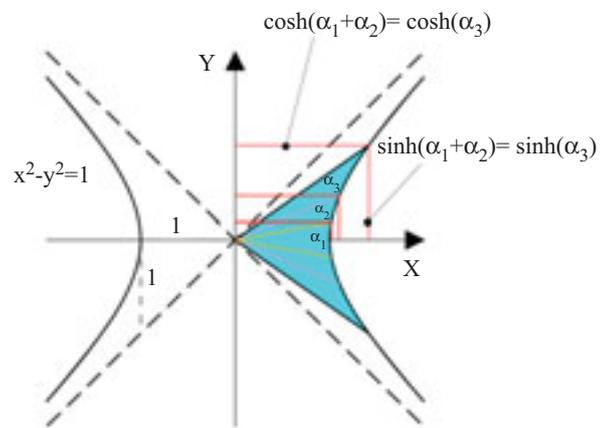

Fig. 5

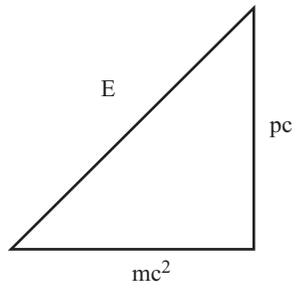

Fig. 6

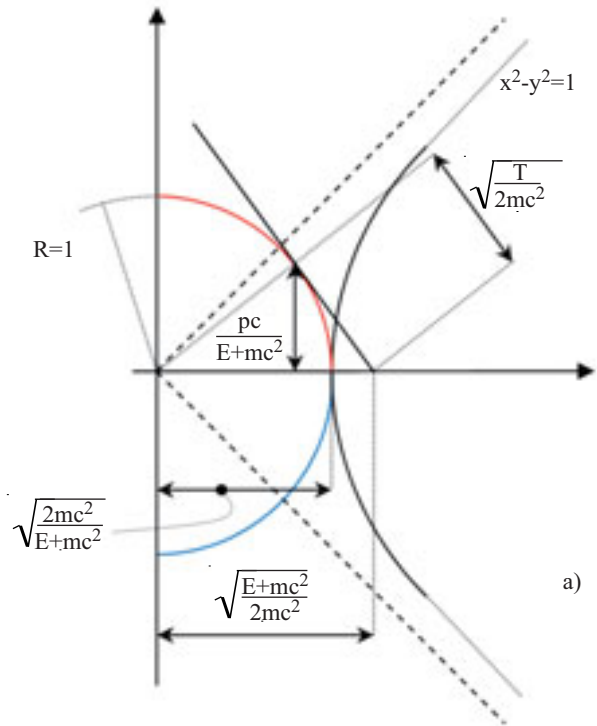

a)

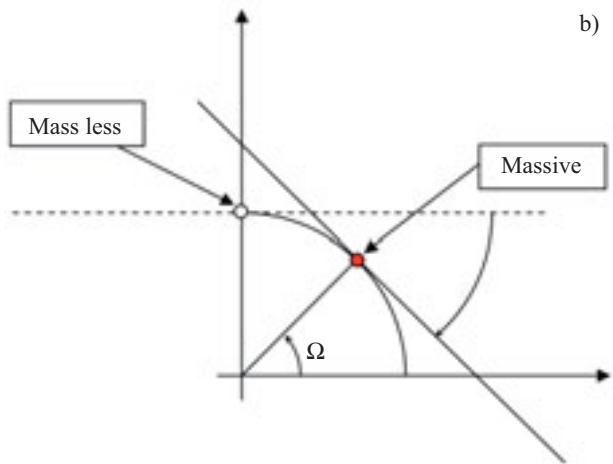

b)

Fig. 7

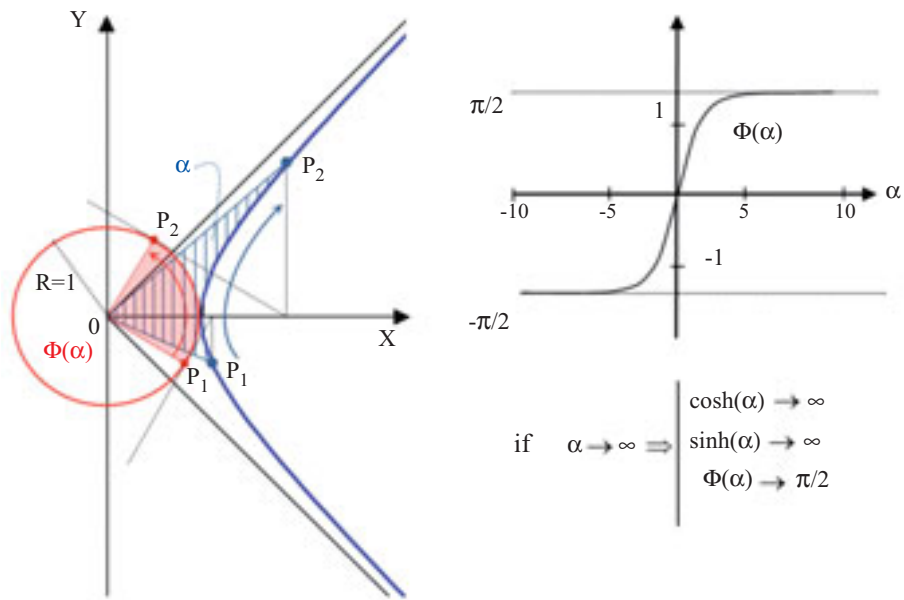

Fig. 8

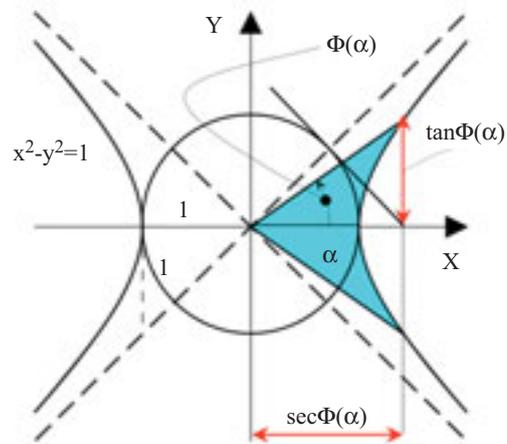

Fig. 9

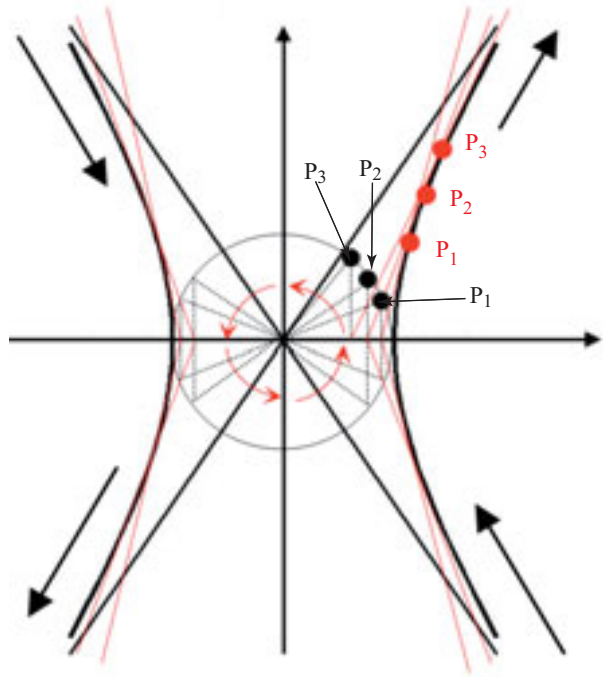

Fig. 10

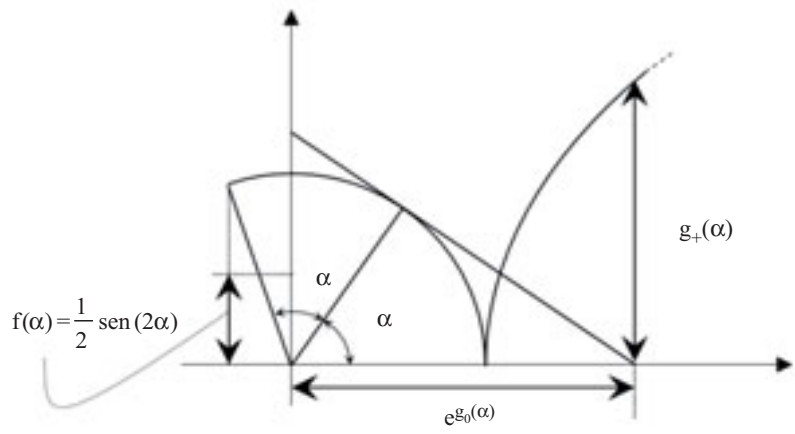

Fig. 11

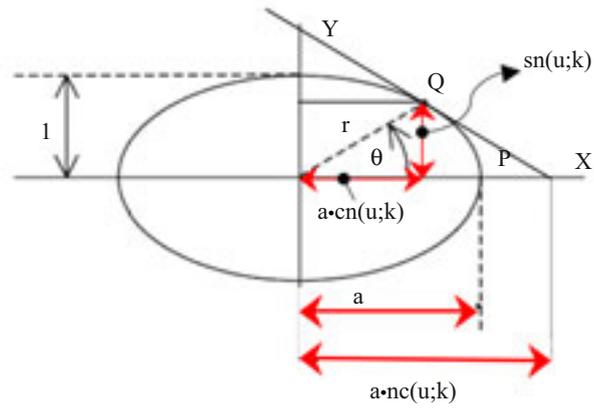

Fig. 12

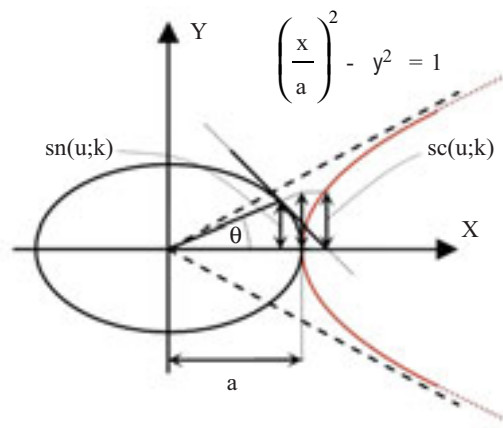

Fig. 13

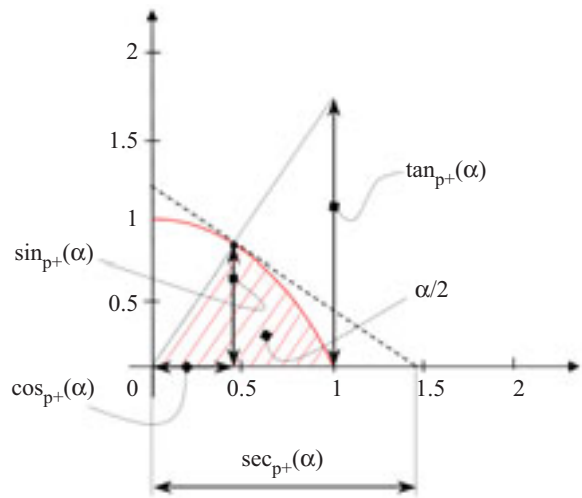

Fig. 14